\documentclass[12]{article}
\usepackage{times}

\newif\ifpdf
\ifx\pdfoutput\undefined
\pdffalse 
\else
\pdfoutput=1 
\pdftrue
\fi

\ifpdf
\usepackage[pdftex]{graphicx}
\else
\usepackage{graphicx}
\fi

\title{Deep Underground Science and Engineering Lab \\
Dark Matter Working Group 2007 White Paper}
\author{D.S.~Akerib and R.J.~Gaitskell, Working Group Co-chairs}
\date{}

\long\def\symbolfootnote[#1]#2{\begingroup%
\def\thefootnote{\fnsymbol{footnote}}\footnote[#1]{#2}\endgroup}
    \def\@fnsymbol#1{\ifcase#1\or *\or \dagger\or \ddagger\or
      \mathchar ``278\or \mathchar ``27B\or \|\or **\or \dagger\dagger
      \or \ddagger\ddagger \else\@ctrerr\fi\relax}

\newcommand{\RnD}{{\small R\&D}}

\newcommand{\CDMS}{{\small CDMS}}

\newcommand{\WIMP}{{\small WIMP}}
\newcommand{\WIMPs}{{\small WIMP}s}

\newcommand{\SUSY}{{\small SUSY}}

\newcommand{\DUSEL}{{\small DUSEL}}

\newcommand{\etal}{et al}	

\begin{document}

\ifpdf
\DeclareGraphicsExtensions{.pdf, .jpg, .tif}
\else
\DeclareGraphicsExtensions{.eps, .jpg}
\fi

\maketitle

\begin{flushright}
\textit{Dec 24, 2007 (final/d)}
\end{flushright}

\bigskip

This whitepaper is the result of discussions and presentations initiated at the DUSEL Town Meeting held in Washington in November 2007. The essential elements of this report are:

\begin{itemize} 

\item The quest to detect dark matter is a science goal of the very highest priority, and is flagship science for DUSEL.
\item The dark matter community presents here a Roadmap for a set of proposals for the Initial Suite of Experiments. The science goals will be reached in two phases of experiments, at the 4850 and 7400 ft levels, respectively.
\item The US is currently the world leader in the search for \WIMP\ dark matter. Constructing \DUSEL\ will ensure that the US will continue its leading role and attract international collaborators to \DUSEL.

\end{itemize}

\bigskip

\section{Science Overview}
\label{overview}

The discovery of dark matter is of fundamental importance to
cosmology, astrophysics, and elementary particle physics~\cite{turner,barish,drell,ostp,epp2010}.\footnote{This section is adapted from the S1 DM working group report~\cite{s1dm}}  A broad
range of observations from the rotation speed of stars in ordinary
galaxies to the gravitational lensing of superclusters tell us that
80--90\% of the matter in the universe is in some new form, different
from ordinary particles, that does not emit or absorb
light. Cosmological observations, especially the Wilkinson 
Anisotropy Probe of the cosmic microwave background radiation, have
provided spectacular confirmation of the astrophysical evidence. The
resulting picture, the so-called ``Standard Cosmology,'' finds that a
quarter of the energy density of the universe is dark matter and most
of the remainder is dark energy. A basic foundation of the model, Big
Bang Nucleosynthesis ({\small BBN}), tells us that at most about 5\%
is made of ordinary matter, or baryons. The solution to this ``dark
matter problem'' may therefore lie in the existence of some new form
of non-baryonic matter. With ideas on these new forms coming from
elementary particle physics, the solution is likely to have broad and
profound implications for cosmology, astrophysics, and fundamental
interactions.  While non-baryonic dark matter is a key component of
the cosmos and the most abundant form of matter in the Universe, so
far it has revealed itself only through gravitational
effects---determining its nature is one of the greatest scientific
issues of our time, making the potential for its discovery and study a key program at DUSEL.

Many potential new forms of matter have been suggested as
dark matter candidates in theories that go beyond the Standard Model of strong and electroweak interactions, but none has yet been produced in the
laboratory.  One possibility is that the dark matter is comprised of
Weakly Interacting Massive Particles, or \WIMPs, that were produced
moments after the Big Bang from collisions of ordinary matter. \WIMPs\
denote a general class of particles characterized by a
mass and annihilation cross section such that they would fall out of
chemical and thermal equilibrium in the early universe at the dark
matter density. Several extensions to the Standard Model lead to
\WIMP\ candidates. One of them is
Supersymmetry (\SUSY), which extends the Standard Model to include a
new set of particles and interactions that solves the gauge hierarchy
problem, leads to a unification of the coupling constants, and is
required by string theory.  The lightest neutral \SUSY\ particle, the
neutralino, is thought to be stable and is a natural dark matter
candidate.  Intriguingly, when \SUSY\ was first developed it was in no
way motivated by the existence of dark matter. This connection could
be a mere coincidence---or a crucial hint that \SUSY\ is responsible
for dark matter.

The possibility that a new class of fundamental particles could be
responsible for the dark matter makes the search for \WIMPs\ in the
galactic halo a very high scientific priority. This will require the use of low-background detectors in a deep underground location, with excellent sensitivity to nuclear recoils and high rejection power for betas and gammas. A direct detection of dark matter in the halo
would be the most definitive way to demonstrate that \WIMPs\ make up the
the missing mass. The study of \WIMP\
candidates in accelerator experiments will play a crucial role in determining the
relic density of these particles, and extensions to the Standard Model
that lead to \WIMP\ candidates are among the primary motivations for
current and next-generation accelerators. The indirect detection of
astrophysical signals due to \WIMP-\WIMP\ self-annihilation may also
provide important clues but in many cases may be difficult to
unambiguously separate from more mundane astrophysical sources. That
leaves direct detection as playing a central role in establishing the
presence of \WIMPs\ in the universe today. Given both the technical
challenge and fundamental importance of direct \WIMP\ detection, it is
vital to have the means to confirm a detection in more than
one type of detector. In addition to giving a critical cross check on
systematic errors that could fake a signal, detection of \WIMPs\ in
multiple nuclei will yield further information about the \WIMP\ mass
and couplings. If eventually \WIMPs\ are discovered, the
ultimate cross check will be confirming their galactic origin by
observing secondary signatures related to the motion of the earth and
solar system, and the velocity distribution of the \WIMPs. Further development of detectors sensitive to the WIMP direction is important to the long-term program to achieve a
state of readiness to vigorously follow up on an initial detection.

{\small US} scientists are in a world-leading position in direct detection by having pioneered the development and deployment of several of the best technologies, and by engaging in an active \RnD\ program that continues this leadership. 

The detection of dark matter is an experimental challenge that requires the development of sophisticated detectors, suppression of radioactive contamination, and---most relevant to this report---siting in deep underground laboratories to shield from cosmic-ray-induced backgrounds. By building the world's premier deep laboratory, together with bringing the ongoing \RnD\ efforts to fruition, the {\small US} will be in a very advantageous position to attract international collaborations and lead the major experiments in this field. Furthermore, given the technical demands of the required scale-up to larger detectors, the establishment of the laboratory and of strong technical and engineering support are essential to the success of these experiments.

\begin{figure}[!h]
\begin{center}
\includegraphics[width=3.2in]{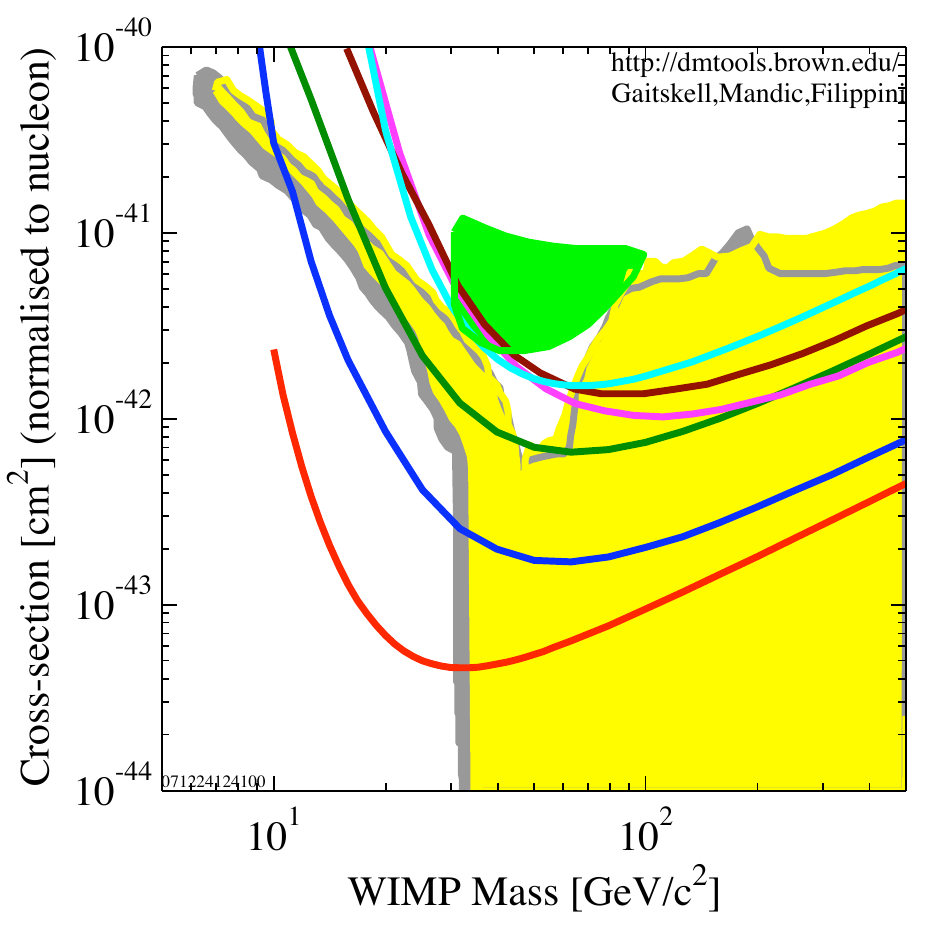}
\includegraphics[width=3.2in]{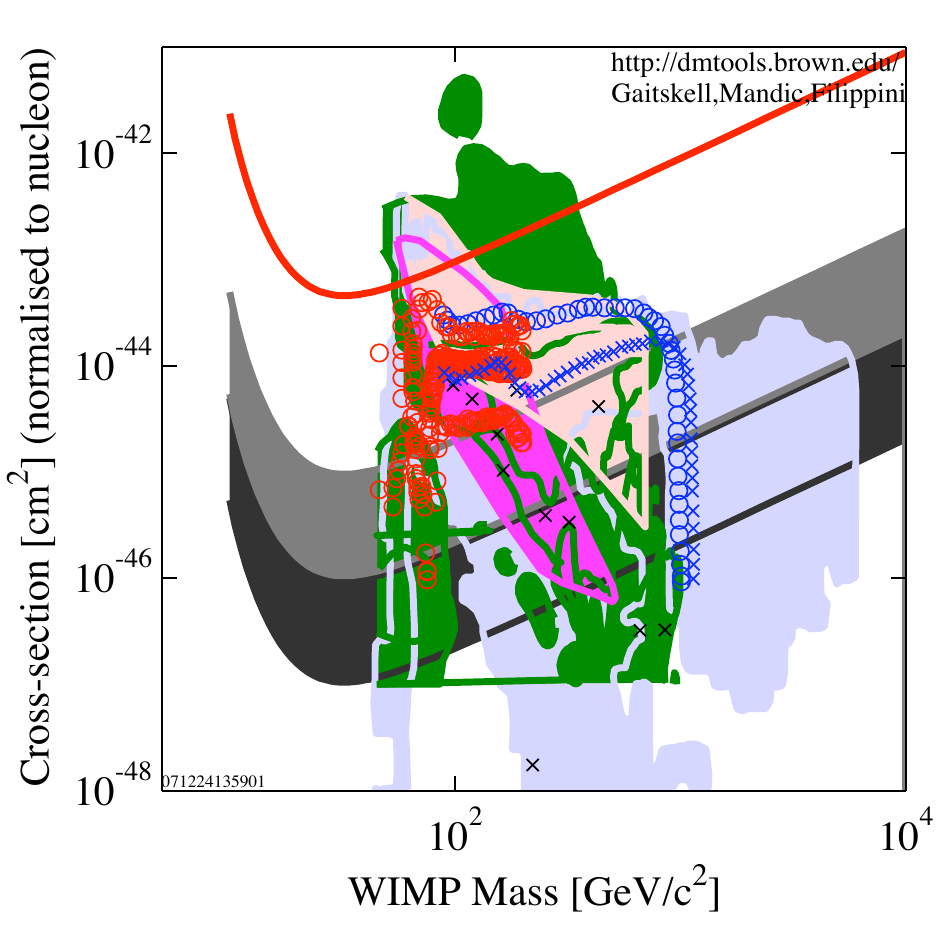}
\end{center}
\caption{\small Plots of the elastic scattering
cross section for spin-independent couplings versus WIMP mass. (a) The
left panel shows the leading experimental results in which the solid
curves represent experimental upper limits from 
the CRESST thermal and scintillation cryogenic
detectors (cyan)~\cite{cresst2005}, 
the
EDELWEISS thermal and ionization cryogenic detectors (dark
red)~\cite{edelweiss},
the WARP two-phase liquid argon
detector (magenta; 55 keV threshold)~\cite{warp}, 
the ZEPLIN-II
two-phase liquid xenon detector (dark green)~\cite{zeplinII}, 
the CDMS-II
athermal-phonons and ionization detectors (blue)~\cite{cdms119}, 
and the XENON-10 two-phase liquid xenon detector (red)~\cite{xenon10}, 
in
all cases assuming the standard halo model~\cite{lewin96} and
nuclear-mass-number $A^2$ scaling. The contested DAMA
annual-modulation claim~\cite{dama} is shown by the green region.  The
yellow and grey regions represent unconstrained Minimal Supersymmetric
Standard Model (MSSM) predictions for low-mass WIMPs that result from
relaxing the GUT-scale unification of gaugino masses~\cite{Bottino03}.
(b) The right panel displays a broad range of models. 
CMSSM models from~\cite{roszkowski2007} are shown in dark green. Well-tempered neutralinos from~\cite{baer2007} are shown in light red. 
Models within the minimal supergravity ({\small
mSUGRA}) framework are shown in 
magenta~\cite{Chattopadhyay04} and light blue~\cite{Baltz04}.
More specific predictions are given by
split Supersymmetry models shown by blue circles and crosses (for
positive and negative values of the $\mu$ parameter,
respectively)~\cite{Giudice04} and red circles~\cite{Pierce04}. The
set of representative post-LEP LHC-benchmark models are shown by black
crosses~\cite{Battaglia03}. Experimental projections are shown as
grey and black regions for the two ISE phases discussed in the Roadmap section of this report, at the 4850- and 7400-foot levels, respectively. The XENON-10 limit is
shown again as a solid red curve for reference with the left panel.}
\label{fig:models_limits}
\end{figure}

The fundamental challenge of the direct detection of \WIMPs\ is based on
elastic scattering between \WIMPs\ in the halo and atomic nuclei in a
terrestrial detector. The energy range for \WIMP-induced nuclear
recoils is of order 10\,keV, an energy range in which the rate
from electromagnetic backgrounds are dominant by many orders of
magnitude. It is therefore essential that \WIMP\ detectors combine
low-radioactivity materials and environments with background rejection
of electron recoil events to keep spurious signals at bay. The dark
matter community has developed a broad range of techniques to address
these challenges. At present the upper limit on the \WIMP-nucleon cross section is below $10^{-43}$\,cm$^2$, which is well into the region of parameter space where SUSY particles could account for the dark matter, as is illustrated in Fig.~\ref{fig:models_limits}. The next 2-3 orders of magnitude represent a particularly rich region of electroweak-scale physics (see~\cite{s1dm} for a full discussion). The combination of what we may learn from astrophysical searches combined with accelerators is truly profound. Thus, as we describe in the following section, the priority for the dark matter community is to press ahead with a pre-DUSEL round of experiments to further explore this parameter space and hone the technological approaches. Then, by building \DUSEL\ and preparing more sensitive dark matter experiments for the ISE we will position ourselves to either follow up and study a dark matter detection in the coming few years or to search completely the remaining parameter space in a discovery mode. 

\section{Priority for initial suite of experiments}

The scientific priority of the DUSEL dark matter ISE is to either detect and confirm an initial signal that can be confidently attributed to \WIMPs, or if a signal (or hint thereof) is observed in the pre-DUSEL era, to move rapidly into an exploitation and study phase. That study would open an era of ``\WIMP'' astronomy that would reduce the astrophysical uncertainties, deduce the nature of the \WIMP-nuclear coupling, measure the \WIMP\ mass, and together with accelerator measurements determine if \WIMPs\ are the full explanation of non-baryonic dark matter. The results of this grand accomplishment would achieve nothing less than confirming our understanding of gravity on a broad range of heretofore-untested distance scales and give us insight into a new form of matter. 

Detection in more than one detector using different target nuclei is essential to gain confidence to eliminate systematic effects, or misidentified backgrounds. Data from additional detectors would permit consistency checks regarding how the cross section scales with nuclear mass and/or spin, and would better exploit kinematic determination of the \WIMP\ mass. Constraining the \WIMP\ properties and couplings will allow for comparison with accelerator-based experiments. Determination of the interaction rate will provide a foundation for designing the follow up experiments to confirm the galactic origin of the signal and the reduction of astrophysical uncertainties.

The current sensitivity of \WIMP\ search experiments is at $\sigma_{SI} \sim 10^{-43}$\,cm$^2$. It is likely that in the pre-DUSEL period the sensitivity level will reach down to about the $10^{-45}$ level. In addition, the LHC may give an indication of new physics that informs \WIMP\ searches, although tight constraints on the elastic cross section are unlikely. This is discussed in the S1 report~\cite{s1dm}. Whether we are still in a search mode or in an exploitation phase, the experiments for the DUSEL ISE will aim for sensitivity of $10^{-46}$\,cm$^2$ -- $10^{-47}$\,cm$^2$/nucleon. Depending on the target nucleus, this goal will require a minimum fiducial mass of $\sim$1 tonne, with 1--10 tonne fiducial likely.  

Improvement factors of about a 1000 in detector-mass scale up and in background reduction, compared to currently operating detectors, appear in reach but will be challenging. Achieving them will rely on maintaining a robust program in the pre-DUSEL period that continues to improve the sensitivity to WIMPs while also providing key technical demonstrations that will inform the ISE. Oriented toward the possibility of galactic-origin confirmation, this would come most definitively from a detector sensitive to the direction of the recoiling nucleus, so R\&D on such technologies should proceed in parallel with scaling up the more conventional calorimetric detectors. While DUSEL planning may be focused on defining and planning for the ISE, we emphasize that the dark matter program requires additional space and support for building and deploying prototype detectors for future generations of experiments. 

In summary, the science priorities are to scan with at least two target types both SD and SI couplings with large enough detectors to discover \WIMPs\ at the $\sigma_{SI} \sim 10^{-46}$\,cm$^2$ level, i.e., on the order of 10 or more events, and a post-discovery phase that exploit multiple targets to determine \WIMP\ mass and couplings (i.e., SI couplings that scale with A$^2$ or SP couplings to odd-n or odd-p nuclei). Following discovery, and continued \RnD\ on directional detection, that approach to galactic confirmation will be defined. To carry out this program the targets will likely chosen from F, Ar, Xe, Ge, Ne, and I. 

Achieving the  sensitivity level will require sufficient depth (overburden) and local shielding to bring unvetoed neutron interactions in the detector material to well below $ 10^{-5}$\,events/kg/day.  A combination of very effective
local shielding and muon vetoes, high radiopurity, event 
discrimination, and accessible well-planned deep laboratory space are
needed to reach this physics-driven goal. The 7400 level at DUSEL will be important to insure that the dark matter detectors can reliably reach to $10^{-46}$\,cm$^2$ -- $10^{-47}$\,cm$^2$/nucleon, and beyond.

\section{Roadmap}

The science priorities for the dark matter field in the next 10+ years are clear. Arriving, immediately, at a {\bf detailed} technical roadmap covering this entire period to pursue those goals is challenging.  For example, our field has seen a rapid evolution of new technologies using noble liquids with a variety of approaches that have a realistic potential to reach the requisite target mass /background levels/WIMP sensitivity at reasonable cost.  At the same time, there are clear cost and engineering challenges to scaling up the well-understood solid-state low temperature germanium detectors. Bubble chamber experiments are advancing in sensitivity, and major scale ups are being considered. Low-pressure gas detectors have the unique capability of sensitivity to  recoil direction, but further \RnD\ is required before major scale ups can be proposed.

The opportunity presented by the MREFC process is highly significant in the context of historical funding profiles for dark matter, and there is a strong argument for organizing our community to take best advantage of it. Fully exploring the science program to achieve a robust discovery will require multiple experiments. To evaluate the candidate experiments, and stage them in a way that is well matched to the laboratory schedule, we propose the following process to arrive at a proposal for the MREFC dark matter ISE. 

A roadmap for the period 2008--2011 (Pre-\DUSEL) has been well studied by the DMSAG committee. However, that committee was not asked to consider the direct impact of the MREFC ISE, as discussed here. The DMSAG recommendations for the field should be followed. However, it is clear that the MREFC ISE selection/submission process (2008--9) will exist in parallel, and will modify the former plans some. For the health of the field it will be important that a strong potential program continues to exist outside the umbrella of the MREFC. Once the funding under MREFC is guaranteed this will clearly lead to some modification of the DMSAG plan, in so far as certain technologies will be being funded directly by the DUSEL program.

For the ISE selection, an {\it ad hoc}, or an existing panel (Homestake {\small PAC} or {\small DMSAG}), or combination thereof, would solicit Letters of Intent for submitting a PDR. These Letters (of order 10 pages) would describe and justify technical plans, sensitivity, cost estimate, make up of collaboration, and schedule, and generally give a preview of plans that would be expanded into a PDR. The community would agree to ``binding arbitration'' of the panel's recommendation for the experiments to be included in the MREFC. We understand that a recommended panel and timetable for this type of process is being studied by the S1 panel. We would recommend that the panel assess submissions in March 2008, to permit enough time for the subsequent MREFC proposals to be written.


Selected programs would make a good faith effort to incorporate new members which would anyway be in their interest given the large scope of the ISE projects. 

Those technologies that are not selected for the MREFC can be maintained at an \RnD\ scale providing the potential for future DUSEL experiments. They will pursue pre-\DUSEL\ goals through the usual peer-review regular program process. 

Realistic cost and schedule estimates are required. To date, we have relied on self-reporting of these profiles for the ISE, and as a field are just at the point where pre-DUSEL experiments are entering detailed stages of review. As the MREFC ISE is defined, it is essential to build in an independent oversight process to assess these estimates. The DMSAG chaired by Hank Sobel reported in 2007 primarily on the pre-DUSEL era with the recommendation for further review in 2009 to prioritize the different approaches. ISE proponents will prepare for detailed reviews that address: 

\begin{itemize}

\item Cost and FTE estimates of progenitor experiment to serve as baseline
\item R\&D needed to finalize detector design and demonstrate sensitivity reach
\item Development required for construction start (subject of S4 proposals but too late to affect a late 2008 MREFC selection process)
\item Estimated cost, schedule, manpower and sensitivity (signal and background) with as detailed justification as possible
\item Estimate of DOE/International component 

\end{itemize}

The evolution of experiments under the ISE would follow a phased construction matched to the two main campuses at \DUSEL, with a milestone review of each experiment after their first phases. For example, using the CTF/Borexino experiment as a model, the MREFC would contain a detailed PDR for the first phase, which would start construction in 2011 for installation at the 4850 level. The second phase, for construction in 2015 at the 7400 level, would be described at a less detailed level in the MREFC and release of funds would require that technical milestones from phase I be reached. This approach has the benefit of mitigating risk and allowing the more expensive larger phase to derive considerable technical benefit from its progenitor. In addition, the larger more sensitive detector requires greater depth for background suppression and is built and deployed on a time frame commensurate with laboratory construction. 

Based on the current understanding of the capability of technologies within the field it seems likely that the Pre-\DUSEL\ experiments (2008--2010) will deploy gross masses of 25--500 kg with a physics reach of $10^{-44}$--$10^{-45}$ cm$^2$. The ISE proposal would contain details of a phased construction of the order of 3-tonne (gross) detectors (reach $10^{-45}$--$10^{-46}$ cm$^2$) at 4850 level (2011--2013) and 10+ tonne (gross) detectors (reach $10^{-46}$--$10^{-47}$ cm$^2$) at the 7400 level (2015--). The MREFC would contain the full PDR for the 4850-level ISE phase, with more limited technical description of the 7400-level ISE phase. Development work under the first phase would be used to establish full feasibility of the second phase.

The cost estimates plus contingency for the second phase experiments would need to be used to define an likely experiment cost envelope for the period 2015-- which would be used in the MREFC. The development of PDR level specification of the second phase experiments would occur in 2012. Technologies not selected for the ISE first phase could be considered for the second phase.


In defining the road map above, given the speed at which some of the technologies have been improving there is the possible risk that the science associated with a reach of $10^{-45}$--$10^{-46}$ cm$^2$ gets done earlier than 2011 at another non-US underground laboratories.
 
A broad summary of estimates relating to proposed experiments are shown in the Table 1.


\begin{table}
\begin{center}
{\small
\begin{tabular}[h]{|c||c|c|c|c||l|}
\hline
Experiment	& Fiducial (Gross) 	&DM  Reach c-s	& Depth (ft Level)		& Proj. 		& Comments including \\
Type       		& Mass [tonne]          		& cm$^{2}$	& / Shielding	& Cost (\$M)		& readiness for PDR\\ 
\hline \hline
LXe TPC 		& 1.5 (3)  		& $0.7\times10^{-46}$	&  4850 w/3-m water			& 10-15		& Costs based on existing technologies \\ 
\hline
LXe TPC 		& 8 (10)  		& $1\times10^{-47}$	&	7400		& 20-35		& \\ 
\hline
\hline
LAr TPC 		& 1 (3)  		& $5\times10^{-46}$	&	4850 w/3-m water		& 7		& Requires reduced $^{39}$Ar \\ 
\hline
LAr TPC 		& 5 (10)  		& $5\times10^{-47}$	& 7400	& 20 & Requires reduced $^{39}$Ar \\ 
\hline
\hline
LAr Single 		& 10 (40)  		& $2\times10^{-47}$ & 7400 & 55 & CLEAN \\ 
\hline
LNe Single 		& 10 (40)  		& $1\times10^{-46}$ & 7400 & 55 & CLEAN \\ 
\hline
\hline
Cryo Ge 		& 0.9 (1)  		& $1\times10^{-46}$ & 7400 & Goal 50 & SuperCDMS - R\&D required for industrialization  \\ 
\hline
\hline
Bubble Ch. 	& 1 (1)  		& $4\times10^{-46}$ & 4850 w/3-m water & 2 &   \\ 
\hline
\hline
LP Xe Gas		&   		&  & &  & CS$_{2}$, CF$_{4}$ promising. At R\&D phase for large mass.  \\ 
\hline
HP Xe Gas		&   		&  & &  & Still at R\&D phase  \\ 
\hline
\hline
\end{tabular}
}
\end{center}
\caption[Experiment Table]{Future candidate experiments for dark matter searches at DUSEL. The Project Cost is equipment and engineering costs, and does not include operations, cavity cost, or basic outfitting. HP - High pressure $>1$ atm. LP - Low pressure $<1$ atm. In the depth column, 4850 and 7400 refer to the depth level in feet of the to primary physics labs at DUSEL}
\label{tbl:experiment_table}
\end{table}

As this process is carried out, the S4 proposals will serve as additional ``markers'' for MREFC-bound bids. However, the current timeline for the MREFC submission itself appears sufficiently rapid that S4-supported work will not directly impact it. However, S4 support will be critical to developing full-scale construction-ready plans for the anticipated MREFC construction start. In some cases, the PDR will detail some options for a technology choice, e.g., readout method for a target type, which could facilitate cooperation in the near term.

The technical roadmap is aligned with the science program goals. The primary science goal is initially extending the sensitivity to WIMPs with spin-independent (SI) couplings since these are expected to dominate in most models. A secondary priority is to push sensitivity for spin-dependent (SD) couplings, which for some nuclei and couplings (xenon and germanium for odd-neutron) are included in the ``SI'' searches; for odd-proton couplings other nuclei (e.g., fluorine or iodine) are required. At present it does not appear that these experiments can match the physics sensitivity reach of SI couplings, which reduces their relative merit, during the period when a WIMP signal has yet to be observed.

An important aspect of the sensitivity goal is to go beyond limit-setting and make a robust discovery. Therefore we have placed significant importance in pursuing at least two experiments that are sensitive to the higher-priority SI couplings.

\section{R\&D needs}

Current \WIMP\ search experiments employ a range of technologies to shield or reject background sources, and these form the basis from which to draw the first dark matter experiments at \DUSEL. These technologies include noble liquids and cryogenic detectors, which together define the forefront of this field, along with the different and unique capabilities of bubble chambers and gaseous detectors. The critical aspects of these technologies required to provide the sensitivity reach of the \DUSEL\ era of dark matter experiments depends on cost/scalability, radiopurity, background rejection, and stability --- as well as sensitivity, which depends primarily on low energy threshold, instrumented fiducial mass, and the properties of the target nucleus. Carrying out the pre-\DUSEL\ experiments which will continue to substantially advance the sensitivity frontier are the primary foundation on which the \DUSEL\ program will be built. In that sense, they are a key {\it development} stage. In addition, specific needs for \RnD\ to ensure a vigorous and flexible program include:

\begin{itemize}

\item Background assay and benchmarking tools (see white paper from that cross-cutting working group 

\item Advanced low-background PMTs and/or alternative scintillation readout schemes

\item Industrialization of cryogenic detector production and production of higher-mass modules

\item Gas detector readout to address head-tail discrimination and higher pressure operation

\end{itemize}

It needs to be established whether \RnD\ costs can be included in the MREFC ISE, and/or appear in DUSEL Laboratory operations costs, or whether they will to continue to come only from the agencies. The funding levels for \RnD\ will need to be increased substantially to ensure the resulting technologies have been demonstrated at a level of proficiency consistent with their deployment at the multi-tonne scale.

\section{Other considerations}

In defining the dark matter ISE within the overall MREFC cost profile, a mitigating factor is the potential for multi-purpose science across other sub-disciplines addressed by the double-beta decay and solar neutrino working groups. Specifically, if large Xe or Ge targets are pursued then added value accrues in the context of the overall ISE if double beta decay science goals can be met; at least a process to ``join forces'' must be considered, although we must be wary of distraction --- it will not serve the science interests of either program if compromises are made in trying to oversell a given approach. 
A 10 tonne Xe dark matter experiment, even containing natural isotopes, would have significant sensitivity to the neutrinoless double beta decay (DBD) mode. The energy resolution of a Xe TPC could be sufficiently good to separate the $2\nu$ and $0\nu$ modes. However, for the unambiguous discrimination of a very weak gamma line from the $0\nu$ mode would require Ba tagging which may be a significant complication of the detector. 
A purchasing program for accumulating target materials over a period of time for multi tonne experiments should be considered, especially if isotopically-enriched material is needed. (e.g., Xe rich in $^{136}$Xe, Ar low in $^{39}$Ar) 
Also, as experiments reach to the $10^{-46}$--$10^{-47}$\,cm$^2$/nucleon level, sensitivity to astrophysical neutrino sources is also of interest. Cross-fertilization of the working groups should be nurtured to explore these possibilities.

\section{E\&O} 

The dark matter community has had a strong record of education and outreach programs that have capitalized on the public's fascination with cosmological questions.  The accessibility of the subject owing to the non-specialists familiarity with gravity presents us with the opportunity to draw them in with the surprising statement that if gravity behaves as we understand it according to the household names of Newton and Einstein, then we have a profound mystery on our hand of not being able to describe or locate most of the matter in the universe, starting right here at home in our own Milky Way galaxy. 

Building on this exciting question, there is also the opportunity to draw in the public with the interesting ``gadgets'' that we are compelled to develop to carry out the search for these ghost-like particles. 

The subject of dark matter searches is a fertile source of material for drawing in, informing and educating the public about a specific profound lack of understanding about nature --- and how we set ourselves about to answer it. It touches on the questions of the origin, formation and fate of the astrophysical structures that appear necessary for our existence. The search for answers is a dynamic illustration of the scientific method itself: how we construct and test hypotheses, what assumptions those hypotheses rest on, and the limitations of  empirical evidence that we will find, be it for or against the existence of \WIMPs. 

The dark matter community at Dusel will be an excellent resource for the E\&O team, and will offer numerous opportunities for visiting K-12 teachers, college teachers, and students of all ages to learn about astro-particle physics. Longer term programs, such as summer teacher training or research experiences for undergraduates from area schools, will also will be an important component of the program. It has also been suggested that university groups consider hosting summer programs in which South Dakota students spend time in summer school programs at the home institutions which can be broadening and formative experiences.\footnote{Frank Calaprice's group at Princeton University has developed and carried out a very successful summer program with Italian high-school students from the Gran Sasso area.}


\end{document}